%% file: template.tex
\documentclass{article}

\usepackage{arxiv}
\usepackage{placeins}
\usepackage[utf8]{inputenc} 
\usepackage[T1]{fontenc}    
\usepackage{hyperref}       
\usepackage{url}            
\usepackage{booktabs}       
\usepackage{amsfonts}       
\usepackage{nicefrac}       
\usepackage{microtype}      
\usepackage{lipsum}
\usepackage{graphicx}
\usepackage{graphicx} 
\usepackage{amsmath}
\usepackage{amsfonts}
\usepackage{pdflscape}
\usepackage{longtable}
\usepackage{rotating}
\usepackage{adjustbox}
\usepackage{colortbl}
\usepackage{xcolor}
\usepackage{amsmath, empheq}
\usepackage{diagbox}
\graphicspath{ {./images/} }
\usepackage{subcaption}

\title{DNA Language Models:  \\
An Assessment of Pre-Training for Fine-Tuning Tasks}

\author{
 Karpinsky Romain \\
  CNRS, LORIA\\
  Nancy, 54000, France   \\
  \texttt{} \\
   \And
 Julien  Mozziconacci \\
  CNRS\\
  \texttt{julien.mozziconacci@mnhn.fr} \\
  \And
 Mickaël Delcey \\
  CNRS; LORIA\\
  Nancy, 54000, France  \\
  \texttt{mickael.delcey@loria.fr} \\
}

\begin{document}
\maketitle
\begin{abstract}

Recent breakthroughs in foundation models and Large Language Models (LLMs) have introduced new opportunities for studying and decoding genomic sequences. Several state-of-the-art approaches, such as DNABERT2, rely on transformer-based architectures, while others, such as ConvNova, still build upon more conventional convolutional models. However, systematic benchmark comparisons across these methods remain scarce. Given that transformer-based models require extensive and costly pretraining, it is crucial to evaluate whether their performance gains justify this overhead. Moreover, LLMs such as DNABERT2 typically rely on Byte Pair Encoding (BPE) tokenization, whose relevance for DNA sequence representation is still debated within the genomics community. In this work, we investigate three key questions: (i) do transformer-based models provide sufficient improvements on fine-tuning tasks upon heavy pretraining, (ii) what is the actual contribution of pretraining in this setting, and (iii) how does BPE tokenization impact performance on genomics-related tasks?

\end{abstract}


\section{Introduction}

Genomics seeks to decode the information embedded in DNA sequences and link it to cells and organisms phenotypes. Advances in high-throughput sequencing have generated unprecedented volumes of genomic data, including full genome sequences, genome polymorphisms and annotation landscapes, which include gene positions, transcription factor binding sites and histone modifications. Leveraging these data requires computational models capable of capturing complex, context-dependent sequence patterns. Machine learning has long supported genome annotation and functional prediction, but recent breakthroughs in deep learning have led to a new era—deep genomics—where neural networks learn directly from raw sequences to predict regulatory elements, interpret non-coding variants, and even design synthetic DNA \cite{routhier2022genomics}.
Early successes relied on convolutional neural networks (CNNs), such as DeepSEA \cite{zhou2015predicting} and Basset \cite{kelley2016basset}, which demonstrated that local sequence motifs could be learned from one-hot encoded DNA. More recently, transformer-based architectures have enriched this landscape and foundation models have emerged for genomic sequences, inspired by large language models (LLMs) in natural language processing. These genomic language models (gLMs), such as DNABERT2 \cite{zhou2023dnabert}, Nucleotide Transformer \cite{dalla2025nucleotide}, and Evo \cite{nguyen2024sequence} aim to capture the genome regulatory logic through self-supervised pretraining on billions of nucleotides. In principle, gLMs could enable zero-shot prediction of variant effects, cross-species annotation transfer, and rational design of regulatory sequences.
However, the non-coding genome poses unique challenges: regulatory signals are sparse, cell-type-specific, and embedded in variable contexts, making it unclear whether generic pretraining objectives confer meaningful biological understanding. Moreover, gLMs introduce substantial computational overhead and sometimes rely on tokenization schemes—such as Byte Pair Encoding (BPE)—whose relevance for DNA remains debated. Despite the growing diversity of architectures, systematic benchmarks comparing transformers to lightweight CNNs, and assessing the impact of tokenization and pretraining, are scarce.

In this work, we address three key questions:

(i) How does BPE tokenization influence model accuracy compared to simpler character-level representations?

(ii) Do transformer-based models deliver significant improvements over convolutional architectures on fine-tuning tasks after heavy pretraining?

(iii) What is the actual contribution of large-scale pretraining to downstream performance in genomics?

To answer these questions, we evaluate DNABERT-2 \cite{zhou2023dnabert}, U-Net \cite{wu2022uiu} , and ConvNova \cite{bo2025revisiting}, under controlled conditions using the Genome Understanding Evaluation (GUE) benchmark, providing critical insights into the trade-offs between model complexity, pretraining requirements, and tokenization strategies.

\section{Models}

In this section, we describe the three architectures evaluated in our study: DNABERT-2, a transformer-based foundation model; U-Net, a classical convolutional architecture adapted from biomedical image segmentation; and ConvNova, a recent CNN-based model specifically designed for genomic sequences.

\subsection{DNABERT-2}

DNABERT-2 is a foundation model for DNA sequences based on the Transformer architecture with 117M parameters. Unlike its predecessor DNABERT, which used overlapping k-mer tokenization, DNABERT-2 adopts Byte Pair Encoding (BPE) with a vocabulary of 4,096 tokens. This choice reduces tokenized sequence length by approximately 5-fold compared to k-mer approaches, substantially improving computational efficiency. The model replaces learned positional embeddings with Attention with Linear Biases (ALiBi), enabling it to handle sequences of arbitrary length without the hard input constraints faced by standard BERT models. To further enhance efficiency, DNABERT-2 incorporates FlashAttention for faster attention computation and uses GEGLU activation functions instead of standard ReLU. 

\subsubsection{Byte Pair Encoding Tokenization}

Byte Pair Encoding (BPE) \cite{shibata1999byte} is a data compression algorithm that has been widely adopted for subword tokenization in natural language processing. The algorithm operates by iteratively identifying and merging the most frequently co-occurring pairs of characters or character sequences in a training corpus. Starting with a base vocabulary of individual characters (in the case of DNA sequences: A, T, C, and G), BPE builds a larger vocabulary by repeatedly finding the most common adjacent pair in the corpus and treating it as a single token. This process continues until a target vocabulary size is reached.

For DNA sequences, the tokenization process can be formalized as follows. Let $\mathcal{V}_0 = \{A, T, C, G\}$ denote the initial vocabulary. At each iteration $k$, the algorithm:
\begin{enumerate}
    \item Scans the entire corpus tokenized with vocabulary $\mathcal{V}_{k-1}$
    \item Identifies the most frequent adjacent pair of tokens $(t_i, t_{i+1})$
    \item Creates a new merged token $t_{new} = t_i \cdot t_{i+1}$
    \item Updates the vocabulary: $\mathcal{V}_k = \mathcal{V}_{k-1} \cup \{t_{new}\}$
    \item Replaces all occurrences of the pair $(t_i, t_{i+1})$ with $t_{new}$ in the corpus
\end{enumerate}

This iterative process generates tokens of variable length that reflect the statistical patterns in the training data. Unlike k-mer tokenization, which uses fixed-length sliding windows with predetermined stride (either overlapping or non-overlapping), BPE adaptively discovers frequent patterns in the data. The resulting vocabulary typically contains both short tokens (individual nucleotides or short motifs) and longer tokens representing common subsequences.

The choice of vocabulary size represents a trade-off: larger vocabularies enable more compression (fewer tokens per sequence) but require more parameters in the embedding layer and may suffer from sparse token frequency distributions during training. In DNABERT-2, empirical evaluation across different vocabulary sizes (from $2^8$ to $2^{15}$) determined that $2^{12} = 4{,}096$ tokens provided the optimal balance between computational efficiency and model performance.

\subsection{U-Net}

U-Net \cite{ronneberger2015u} is a convolutional neural network architecture originally designed for biomedical image segmentation. It features a symmetric U-shaped structure consisting of a contracting path (encoder) and an expanding path (decoder). We used an implementation \cite{wu2022uiu} in which the contracting path applies repeated 3×3 convolutions followed by ReLU activation and 2×2 max pooling operations, doubling the number of feature channels at each downsampling step to capture increasingly abstract context. The expanding path performs 2×2 up-convolutions that halve the number of channels, concatenates the correspondingly cropped feature maps from the contracting path (cropping compensates for border pixel loss in unpadded convolutions), and applies two 3×3 convolutions with ReLU activation. This design allows high-resolution features from the contracting path to be combined with upsampled context from the decoder, enabling precise localization. 

\subsection{ConvNova}

ConvNova \cite{bo2025revisiting} is a CNN-based architecture for DNA modeling with parameter counts ranging from 1.7M to 27.4M depending on configuration. The model's core innovation is the Gated Convolution Block (GCB) with a dual-branch structure: one branch extracts features through convolution and GELU activation, while the other generates gating signals via convolution and sigmoid activation, enabling dynamic feature selection at each spatial location. Each GCB uses dilated convolutions with kernel size 9 and progressively increasing dilation rates [1, 1, $d$, $d^2$, $d^3$, ...] to expand the receptive field without downsampling.

\section{Training Scheme}

\subsection{Pre-Training dataset}

To train our models, we assembled the same pretraining dataset of \cite{zhou2023dnabert} with 135 distinct species distributed across major biological groups, including mammals, invertebrates, fungi, protozoa, bacteria, and other vertebrates. The collection amounts to roughly 32.5 billion nucleotides, representing an order of magnitude increase compared to the human reference genome alone. This heterogeneous dataset provides a rich spectrum of evolutionary variation, designed to encourage the model to capture cross-species regularities and transferable genomic features rather than overfitting to a single organism.

\subsection{Masked Language Modelling}

All models are pretrained with a \textit{masked language modeling} (MLM) objective applied at the token level, either using byte-pair encoding (BPE) tokens or nucleotides. Let 
\[
t = (t_1, \dots, t_m)
\] 
denote the tokenized input sequence, with vocabulary $\mathcal{V}$. During pretraining, we randomly sample a subset of token positions $\mathcal{M} \subset \{1,\dots,m\}$ to be masked, following the standard BERT masking procedure (i.e., replacing each masked token with the special \texttt{[MASK]} token 80\% of the time, with a random token 10\% of the time, and keeping it unchanged 10\% of the time). The resulting corrupted sequence is denoted $t^{(\text{mask})}$.  

For a fixed masked sequence $t^{(\text{mask})}$, the model defines a conditional probability distribution over the vocabulary:
\begin{equation}
t \mapsto p_\theta(t \mid t^{(\text{mask})})~t\in \mathcal{V}, 
\end{equation}
which assigns a probability to each candidate token at every masked position.  

The MLM pretraining loss is given by the average negative log-likelihood of the true tokens at the masked positions:
\begin{equation}
\mathcal{L}_{\mathrm{MLM}}(\theta) 
= - \frac{1}{|\mathcal{M}|} \sum_{i \in \mathcal{M}} \log p_\theta\!\big(t_i \mid t^{(\text{mask})}\big).
\end{equation}

This objective encourages the model to recover the original tokens from their masked context. Unless otherwise specified, all models are pretrained using this same procedure, ensuring that differences in performance are attributable to model architecture rather than pretraining strategy.

All models are trained for 30 epochs using the AdamW optimizer \cite{loshchilov2017decoupled}. We use a one-cycle learning-rate schedule~\cite{smith2017cyclical}, with a maximum learning rate of $10^{-3}$, and a batch size of 512 regardless of the tokenizer.

\subsection{Fine-tuning on Genome Understanding Evaluation benchmark}

For downstream evaluation, we adopt the Genome Understanding Evaluation benchmark (GUE), following the same protocol used in DNABERT-2. GUE integrates 28 datasets across seven representative tasks, including human promoter and core promoter identification, transcription factor (TF) binding prediction in human and mouse, human splice site detection, yeast epigenetic mark prediction, and SARS-CoV-2 variant classification. Sequence lengths range from 70 to 1000 bases, and the datasets are standardized with predefined train/validation/test partitions and unified metrics. By using the same benchmark as DNABERT-2, we ensure that our results are directly comparable and that improvements can be attributed to the modeling choices rather than dataset artifacts.

\begin{figure}[h]
    \centering
    \includegraphics[width=\linewidth]{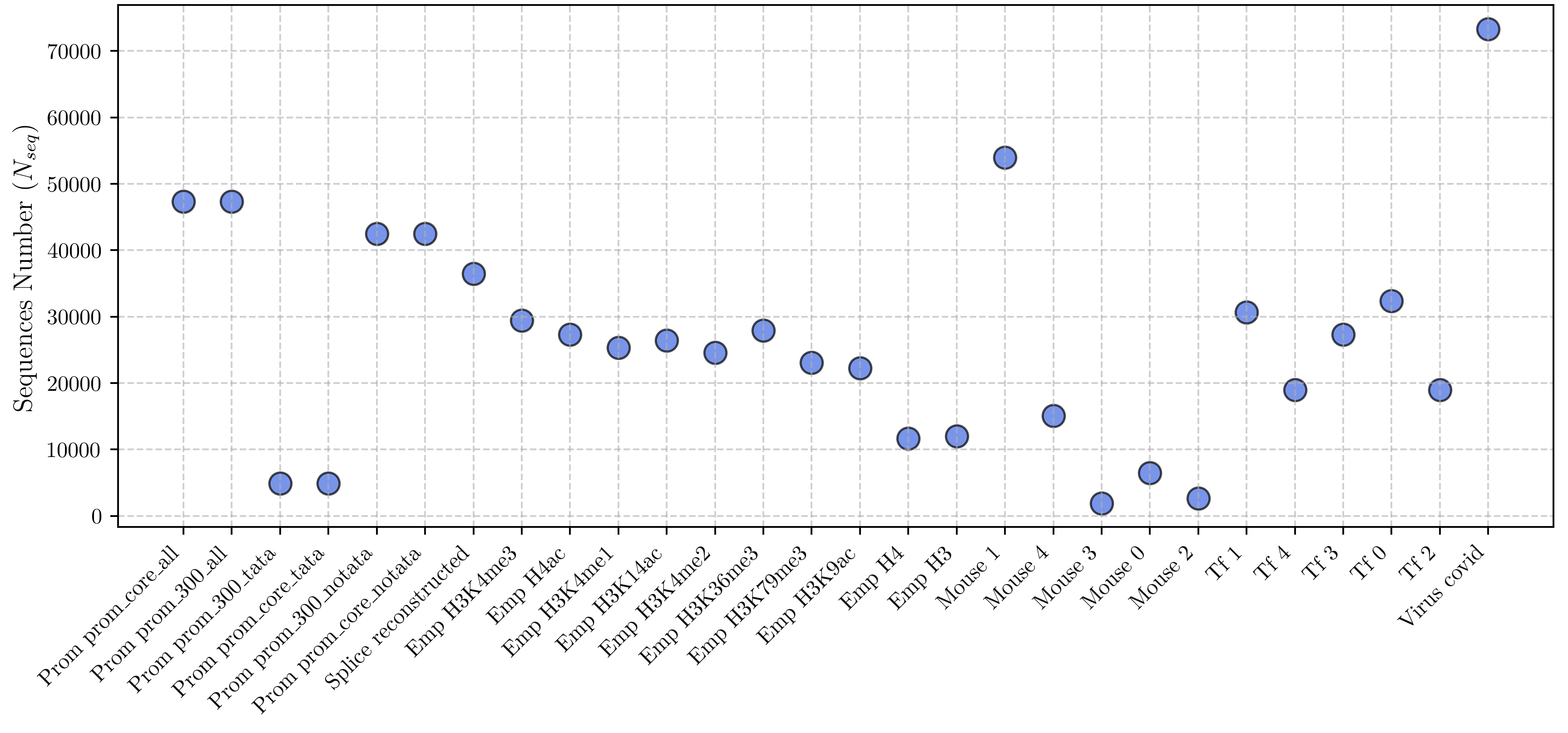}
    \caption{Number of sequences ($N_{\mathrm{seq}}$) for each dataset in the GUE benchmark. The datasets span a wide range of sizes, from a few thousand sequences in certain promoter and mouse tasks to more than 70,000 sequences in the virus classification dataset.}
    \label{fig:gue_sizes}
\end{figure}

To better contextualize the diversity of the GUE, we provide descriptive statistics of the datasets in terms of sample size and sequence length. As shown in Figure~\ref{fig:gue_sizes}, the number of sequences varies substantially across tasks, ranging from a few thousand samples for certain promoter or mouse TF datasets to more than 70,000 sequences for the virus classification benchmark. This variability reflects the heterogeneous nature of genomic prediction tasks, spanning both data-limited settings and large-scale classification problems.

\begin{figure}[h]
    \centering
    \includegraphics[width=\linewidth]{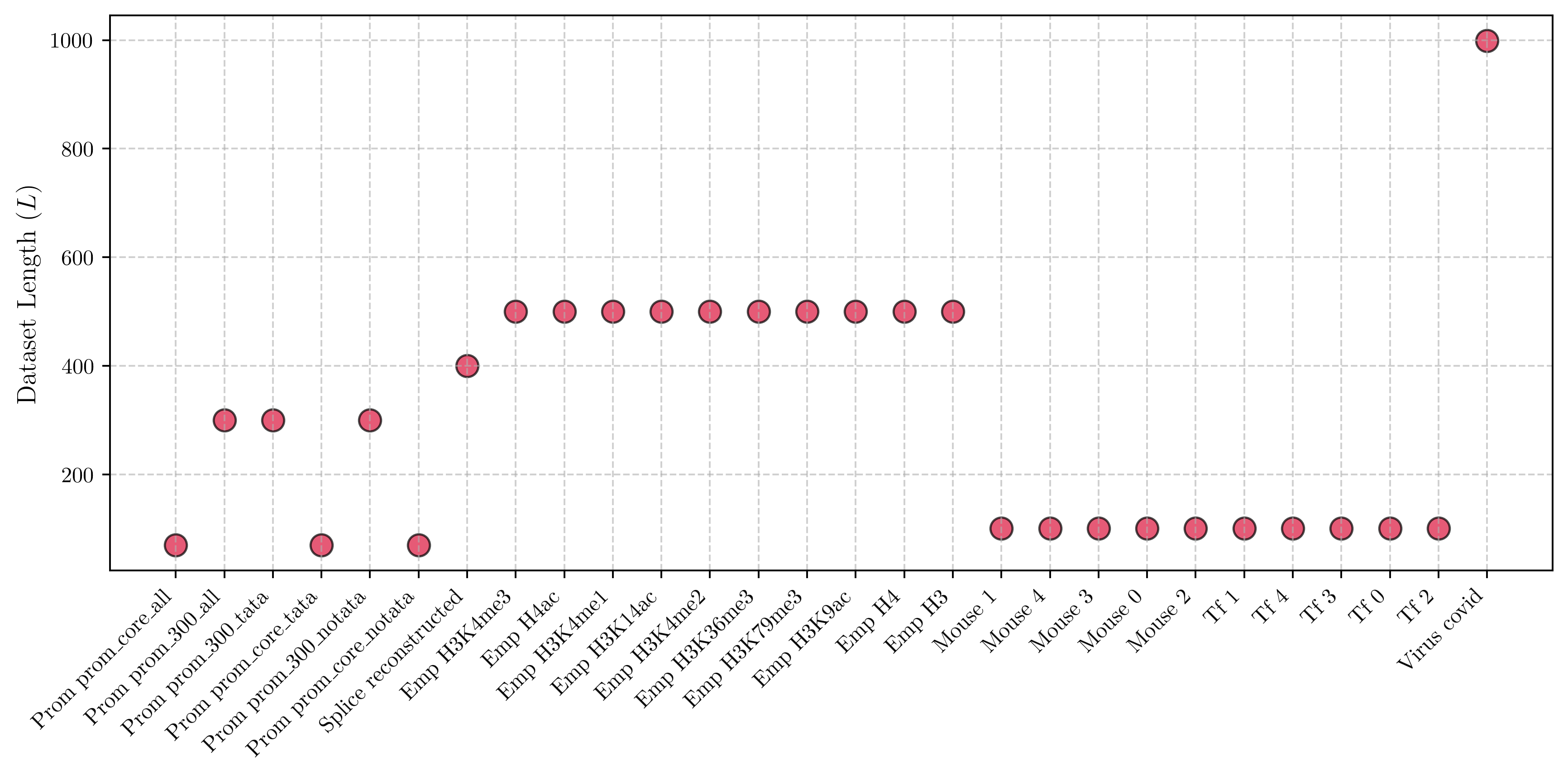}
    \caption{Sequence length ($L$) across datasets in the GUE benchmark. Most tasks involve short sequences (70--100 bases), while others such as EMP tasks use longer sequences ($\sim$500 bases), and the virus classification dataset contains sequences up to 1000 bases.}
    \label{fig:gue_lengths}
\end{figure}

Complementarily, Figure~\ref{fig:gue_lengths} shows the distribution of sequence lengths across datasets. Most tasks operate on relatively short sequences (70--100 bases), while others such as epigenetic mark prediction in the EMP benchmark use longer sequences (around 500 bases), and the virus classification task reaches sequence lengths of up to 1000 bases. This diversity in input length further highlights the challenge of designing models that generalize across tasks with different receptive field requirements.

Together, these statistics emphasize that the GUE covers a broad range of dataset sizes and sequence lengths, providing a comprehensive testbed for assessing the robustness of different modeling approaches under varied genomic regimes.
This heterogeneity in dataset characteristics should be taken into account when interpreting the comparative performance of models across benchmark families.

\FloatBarrier
\section{Effect of Byte Pair Encoding}

To assess the impact of Byte Pair Encoding (BPE) on lightweight architectures, we evaluated U-Net and ConvNova across all GUE benchmark families using both nucleotide-level and BPE tokenization. Results are reported in Tables 1--6, with a global summary in Table 7.
Our key finding is that BPE consistently degrades performance for U-Net and ConvNova, except for the virus classification task, where slight improvements are observed.

\subsection{EMP Benchmark}
The EMP benchmark (10 epigenetic mark prediction tasks) shows a marked performance drop for both U-Net and ConvNova when using BPE (Table 1). \textit{H3K4me2} (U-Net: $-10.96$) and \textit{H3K79me3} (ConvNova: $-8.65$) exhibit the largest losses. U-Net's mean MCC (Table 7) decreases from 47.07 to 35.15 ($-11.92$), while ConvNova's drops from 55.03 to 52.40 ($-2.63$).

\subsection{Mouse Transcription Factor Binding}
For the five mouse transcription factor binding datasets (Table 2), BPE severely impacts U-Net, with MCC scores collapsing in tasks 2 and 4 (e.g., \textit{Mouse-2}: $73.89 \rightarrow 19.97$).
ConvNova also suffers, though less dramatically (mean MCC in Table 7: $67.74 \rightarrow 47.57$).

\subsection{Promoter Identification}
The promoter benchmark (6 tasks, Table 3) reveals task-dependent sensitivity to BPE.
U-Net's performance remains quite good for \textit{prom\_300\_notata} but drops sharply for \textit{prom\_300\_tata} ($82.66 \rightarrow 27.20$) and \textit{prom\_core\_tata} ($84.91 \rightarrow 22.44$).
ConvNova shows losses that overall mirror the losses observed for U-Net.

\subsection{Splice-Site Detection}
Splice-site prediction (Table 4) is the most affected by BPE:
U-Net's MCC plummets from 86.16 to 22.91 ($-63.25$), while ConvNova's decreases from 84.64 to 65.66 ($-18.98$).
This suggests that BPE disrupts local motif detection critical for splice-site identification.

\subsection{Human Transcription Factor Binding}
For human transcription factor binding (5 tasks, Table 5), BPE uniformly reduces performance for both architectures.
U-Net's mean MCC (Table 7) drops from 42.15 to 24.05 ($-18.10$), with the largest loss in task 2 ($50.44 \rightarrow 19.92$).
ConvNova's mean MCC falls from 44.37 to 31.47 ($-12.90$).

\subsection{Virus Classification}
In contrast to other benchmarks, virus classification (Table 6) shows a \textbf{slight performance gain} with BPE:
U-Net improves from 65.99 to 68.07 ($+2.08$), and ConvNova from 68.68 to 70.98 ($+2.30$).
This exception may reflect the longer sequence lengths (up to 1000 bases) in this task, where BPE's compression offsets its disruptive effects on local patterns.

\subsection{Global Summary}
Table 7 aggregates results across all 28 datasets. BPE degrades performance for U-Net and ConvNova in 5 out of 6 benchmark families, with the largest losses in splice-site detection (U-Net: $-63.25$) and mouse TF binding (U-Net: $-35.20$).
The only consistent exception is virus classification, where BPE provides a modest improvement.
A possible interpretation is that Lightweight CNNs lack the representational capacity to leverage BPE's variable-length tokens, which disrupt local motif detection—critical for tasks like splice-site prediction or TF binding.
In contrast, DNABERT-2's self-attention mechanism can adapt to BPE's abstraction, explaining its superior performance with this tokenization scheme.

\input{results_bpe/EMP_table}

\input{results_bpe/mouse_table}

\input{results_bpe/prom_table}

\input{results_bpe/splice_table}

\input{results_bpe/tf_table}

\input{results_bpe/virus_table}

\begin{table}[h]
\centering
\caption{Summary of the impact of BPE tokenization across benchmark families. Each column reports the mean MCC ($\times 100$) over datasets belonging to the corresponding benchmark family. Overall mean is averaged over all 28 datasets.}
\label{tab:summary_bpe_full}
\begin{tabular}{lccccccc}
\hline
Model / Tokenization & EMP & Splice & TF & Mouse & Prom & Virus & Overall \\
\hline
U-Net              & 47.07 & 86.16 & 42.15 & 59.03 & 81.70 & 65.99 & 57.82 \\
U-Net (BPE)        & 35.15 & 22.91 & 24.05 & 23.83 & 47.13 & 68.07 & 34.45 \\
ConvNova           & 55.03 & 84.64 & 44.37 & 67.74 & 74.33 & 68.68 & 61.08 \\
ConvNova (BPE)     & 52.40 & 65.66 & 31.47 & 47.57 & 55.87 & 70.98 & 49.68 \\
\hline
\end{tabular}
\end{table}
\FloatBarrier
\section{Effect of Pretraining}

We evaluate the contribution of pretraining by comparing three conditions: multi-species pretraining, human-only pretraining, and no pretraining (training from scratch). Results are reported in Tables 8--13, with a global summary in Table 14.
Our key finding here is that pretraining is critical for DNABERT-2 but has limited impact on U-Net and ConvNova, whose performance remains stable across training conditions.

\subsection{EMP Benchmark}
For the EMP benchmark (Table 8), DNABERT-2 shows a strong dependence on pretraining, with mean MCC (Table 14) dropping from 55.98 (pretrained) to 42.27 (no pretraining, $-13.71$).
In contrast, U-Net's mean MCC remains nearly unchanged between pretrained and no pretraining conditions (47.07 to 47.11, $\Delta<0.1$), while ConvNova shows marginal variation (55.03 to 54.98, $\Delta=-0.05$).

\subsection{Mouse Transcription Factor Binding}
In the mouse TF binding benchmark (Table 9), DNABERT-2's mean MCC (Table 14) decreases from 68.40 to 50.19 without pretraining ($-18.21$).
U-Net and ConvNova, however, exhibit minimal sensitivity: U-Net's mean MCC ranges from 59.03 to 61.23 ($\Delta<2.2$), and ConvNova's from 67.74 to 68.68 ($\Delta<1$).

\subsection{Promoter Identification}
For promoter tasks (Table 10, 14), DNABERT-2's performance drops from 77.37 to 54.29 without pretraining ($-23.08$).
U-Net and ConvNova again show robustness: U-Net's mean MCC varies by less than 1 point (81.70 to 81.46), and ConvNova's from 74.33 to 75.11 ($\Delta<1$).

\subsection{Splice-Site Detection}
Splice-site prediction (Table 11) reveals a similar pattern: all architectures and learning scheme perform well except DNABERT-2 without pretraining.
DNABERT-2's MCC decreases from 84.99 to 55.53 without pretraining ($\Delta=-29.46$), but U-Net and ConvNova maintain high performance (84.77--87.54 for U-Net, 84.64--87.03 for ConvNova).

\subsection{Human Transcription Factor Binding}
For human TF binding (Table 12), DNABERT-2's mean MCC  (Table 14) falls from 70.11 to 32.88 without pretraining ($\Delta=-37.23$).
U-Net and ConvNova show only minor fluctuations: U-Net's mean MCC ranges from 40.89 to 43.89 ($\Delta<3$), and ConvNova's from 44.21 to 47.22 ($\Delta<3$).

\subsection{Virus Classification}
In virus classification (Table 13), pretraining has limited impact even for DNABERT-2, with MCC decreasing only from 71.02 to 67.90 without pretraining ($-3.12$).
U-Net and ConvNova perform consistently across all conditions (65.02--66.67 for U-Net, 67.87--68.68 for ConvNova).

\subsection{Global Summary}
Table 14 aggregates results across all 28 datasets.
Pretraining is essential for DNABERT-2, which shows an overall mean MCC drop of $-20.91$ without pretraining (66.88 to 45.97).
In contrast, U-Net and ConvNova are largely unaffected: U-Net's mean MCC varies by less than 1 point across conditions (57.45--57.93), and ConvNova's by only $+0.22$ (61.08 to 61.30).
Taken together, these results suggest that DNABERT-2's transformer architecture relies on large-scale pretraining to capture complex, non-local dependencies in genomic sequences. Lightweight CNNs, however, can achieve competitive performance with sufficient labeled data during fine-tuning, as their local kernels effectively capture short-range motifs without requiring extensive pretraining.

\input{results_pretraining/EMP_table}

\input{results_pretraining/mouse_table}

\input{results_pretraining/prom_table}

\input{results_pretraining/splice_table}

\input{results_pretraining/tf_table}

\input{results_pretraining/virus_table}

\begin{table}[h]
\centering
\caption{Summary of downstream performance across training conditions and benchmark families. Each column reports the mean MCC ($\times 100$) over datasets in the corresponding benchmark family. Overall mean is averaged over all 28 datasets.}
\label{tab:summary_pretraining_full}
\begin{tabular}{lccccccc}
\hline
Model / Setting & EMP & Splice & TF & Mouse & Prom & Virus & Overall \\
\hline
DNABERT-2                         & 55.98 & 84.99 & 70.11 & 68.40 & 77.37 & 71.02 & 66.88 \\
DNABERT-2 (no pretraining)        & 42.27 & 55.53 & 32.88 & 50.19 & 54.29 & 67.90 & 45.97 \\
U-Net (pretrained)                & 47.07 & 86.16 & 42.15 & 59.03 & 81.70 & 65.99 & 57.82 \\
U-Net (pretrained-human only)     & 46.85 & 87.54 & 40.89 & 61.23 & 81.46 & 66.67 & 57.93 \\
U-Net (no pretraining)            & 47.11 & 84.77 & 43.89 & 56.23 & 81.20 & 65.02 & 57.45 \\
ConvNova (pretrained)             & 55.03 & 84.64 & 44.37 & 67.74 & 74.33 & 68.68 & 61.08 \\
ConvNova (pretrained-human only)  & 52.59 & 87.03 & 44.21 & 68.68 & 75.11 & 67.87 & 60.57 \\
ConvNova (no pretraining)         & 54.98 & 86.44 & 47.22 & 66.31 & 74.09 & 67.89 & 61.30 \\
\hline
\end{tabular}
\end{table}

\FloatBarrier
\section{Conclusion}

Our comparative analysis of transformer-based and convolutional architectures for genomic modeling highlights important trade-offs between model complexity, tokenization strategies, and pretraining requirements. Across most benchmark families considered here, Byte Pair Encoding (BPE) degrades the performance of lightweight convolutional architectures such as U-Net and ConvNova, with particularly large drops on splice-site, mouse, promoter, and transcription factor binding tasks. At the same time, the effect is not strictly universal, as isolated exceptions remain on virus classification and on a small number of individual datasets. Overall, however, the evidence indicates that BPE is generally not well suited to these lightweight architectures, likely because they do not have sufficient representational capacity to exploit variable-length tokenization effectively.

More broadly, our results reveal a clear distinction between transformer-based and convolutional approaches. DNABERT-2 benefits strongly from large-scale pretraining and achieves the best overall performance on several benchmark families, especially transcription factor binding prediction. In contrast, U-Net and ConvNova remain competitive even without pretraining, and their performance varies only marginally across pretraining strategies. This pattern holds not only on EMP and splice-site prediction, but also on mouse and promoter datasets, where lightweight convolutional models often match or exceed the performance of the pretrained transformer. Although task-specific exceptions remain, the dominant trend is that the benefit of pretraining is far larger for DNABERT-2 than for the smaller convolutional models.

The variability observed across benchmark families can largely be explained by differences in dataset size and sequence length. Tasks involving short sequences and limited contextual dependencies (e.g., transcription factor binding or splice-site prediction) favor character-level representations and lightweight convolutional models, leading to a strong negative impact of BPE. In contrast, tasks involving longer sequences and larger datasets (e.g., virus classification) can benefit from tokenization and exhibit reduced dependence on pretraining. These results highlight that both tokenization strategies and pretraining effectiveness are strongly task-dependent, and are influenced by the underlying data regime.

Finally, when juxtaposed with recent evaluations of genomic language models \cite{tang2025evaluating}, a consistent theme emerges: pretraining on whole genomes does not yet provide a universal advantage. Both studies suggest that pretrained transformer representations offer limited gains without fine-tuning, while lightweight CNNs remain competitive across a wide range of tasks at a fraction of the computational cost. Taken together, these observations argue for a pragmatic strategy: reserve large pretrained transformers for applications that demonstrably benefit from their capacity, while relying on efficient convolutional architectures for motif-centric problems or resource-constrained settings. Future work should focus on more systematic benchmarks, improved interpretability, and domain-informed pretraining strategies to clarify when and why foundation models truly outperform simpler alternatives.

\clearpage

\bibliographystyle{abbrv}
\bibliography{references}

\end{document}

%% file: results_bpe/EMP_table.tex
\begin{table}[!h]

\centering 
\begin{adjustbox}{max width=\textwidth} 
\begin{tabular}{|l|c|c|c|c|c|c|c|c|c|c|}
\hline
\diagbox[width=12em]{Model}{Dataset} & H3 & H3K14ac & H3K36me3 & H3K4me1 & H3K4me2 & H3K4me3 & H3K79me3 & H3K9ac & H4 & H4ac \\
\hline
DNABERT-2 & {\cellcolor[HTML]{B8122A}} \color[HTML]{F1F1F1} 78.27 & {\cellcolor[HTML]{EF886B}} \color[HTML]{F1F1F1} 52.57 & {\cellcolor[HTML]{B40426}} \color[HTML]{F1F1F1} 56.88 & {\cellcolor[HTML]{B40426}} \color[HTML]{F1F1F1} 50.52 & {\cellcolor[HTML]{CBD8EE}} \color[HTML]{000000} 31.13 & {\cellcolor[HTML]{F5C2AA}} \color[HTML]{000000} 36.27 & {\cellcolor[HTML]{B40426}} \color[HTML]{F1F1F1} 67.39 & {\cellcolor[HTML]{CF453C}} \color[HTML]{F1F1F1} 55.63 & {\cellcolor[HTML]{B40426}} \color[HTML]{F1F1F1} 80.71 & {\cellcolor[HTML]{EC7F63}} \color[HTML]{F1F1F1} 50.43 \\
\hline
U-Net & {\cellcolor[HTML]{F49A7B}} \color[HTML]{000000} 71.88 & {\cellcolor[HTML]{A5C3FE}} \color[HTML]{000000} 35.22 & {\cellcolor[HTML]{DBDCDE}} \color[HTML]{000000} 46.74 & {\cellcolor[HTML]{D3DBE7}} \color[HTML]{000000} 35.79 & {\cellcolor[HTML]{82A6FB}} \color[HTML]{F1F1F1} 26.04 & {\cellcolor[HTML]{94B6FF}} \color[HTML]{000000} 24.50 & {\cellcolor[HTML]{F6A586}} \color[HTML]{000000} 62.26 & {\cellcolor[HTML]{F29072}} \color[HTML]{F1F1F1} 53.14 & {\cellcolor[HTML]{DA5A49}} \color[HTML]{F1F1F1} 77.61 & {\cellcolor[HTML]{8CAFFE}} \color[HTML]{000000} 37.55 \\
\hline
U-Net-BPE & {\cellcolor[HTML]{4C66D6}} \color[HTML]{F1F1F1} 53.52 & {\cellcolor[HTML]{3B4CC0}} \color[HTML]{F1F1F1} 23.55 & {\cellcolor[HTML]{3B4CC0}} \color[HTML]{F1F1F1} 36.76 & {\cellcolor[HTML]{3B4CC0}} \color[HTML]{F1F1F1} 23.06 & {\cellcolor[HTML]{465ECF}} \color[HTML]{F1F1F1} 21.69 & {\cellcolor[HTML]{485FD1}} \color[HTML]{F1F1F1} 17.03 & {\cellcolor[HTML]{3B4CC0}} \color[HTML]{F1F1F1} 49.00 & {\cellcolor[HTML]{3E51C5}} \color[HTML]{F1F1F1} 40.76 & {\cellcolor[HTML]{3B4CC0}} \color[HTML]{F1F1F1} 54.21 & {\cellcolor[HTML]{3B4CC0}} \color[HTML]{F1F1F1} 31.88 \\
\hline
ConvNova & {\cellcolor[HTML]{EA7B60}} \color[HTML]{F1F1F1} 73.77 & {\cellcolor[HTML]{F7AA8C}} \color[HTML]{000000} 49.42 & {\cellcolor[HTML]{D75445}} \color[HTML]{F1F1F1} 54.83 & {\cellcolor[HTML]{F4987A}} \color[HTML]{000000} 43.67 & {\cellcolor[HTML]{DADCE0}} \color[HTML]{000000} 32.41 & {\cellcolor[HTML]{BE242E}} \color[HTML]{F1F1F1} 47.42 & {\cellcolor[HTML]{C43032}} \color[HTML]{F1F1F1} 66.54 & {\cellcolor[HTML]{C83836}} \color[HTML]{F1F1F1} 55.92 & {\cellcolor[HTML]{E57058}} \color[HTML]{F1F1F1} 76.56 & {\cellcolor[HTML]{F08B6E}} \color[HTML]{F1F1F1} 49.80 \\
\hline
ConvNova-BPE & {\cellcolor[HTML]{7EA1FA}} \color[HTML]{F1F1F1} 57.48 & {\cellcolor[HTML]{D65244}} \color[HTML]{F1F1F1} 56.55 & {\cellcolor[HTML]{E46E56}} \color[HTML]{F1F1F1} 53.75 & {\cellcolor[HTML]{F7B99E}} \color[HTML]{000000} 41.29 & {\cellcolor[HTML]{B40426}} \color[HTML]{F1F1F1} 44.53 & {\cellcolor[HTML]{C53334}} \color[HTML]{F1F1F1} 46.78 & {\cellcolor[HTML]{D8DCE2}} \color[HTML]{000000} 57.89 & {\cellcolor[HTML]{F59F80}} \color[HTML]{000000} 52.56 & {\cellcolor[HTML]{97B8FF}} \color[HTML]{000000} 61.60 & {\cellcolor[HTML]{E26952}} \color[HTML]{F1F1F1} 51.53 \\
\hline
\end{tabular}
\end{adjustbox}
\vspace{0.1cm}
\caption{MCC scores ($\times 100$) on the EMP benchmark for DNABERT-2, U-Net, and ConvNova, comparing standard nucleotide-level tokenization with Byte Pair Encoding (BPE). The EMP benchmark comprises 10 epigenetic mark prediction datasets. Higher values indicate better performance.}

\label{tab:EMP_mcc_scores_bpe}
\end{table}

%% file: results_bpe/mouse_table.tex
\begin{table}[!h]
\label{tab:mouse_mcc_scores}
\centering 
\begin{adjustbox}{max width=\textwidth} 
\begin{tabular}{|l|c|c|c|c|c|c|}
\hline
\diagbox[width=12em]{Model}{Dataset} & 0 & 1 & 2 & 3 & 4 \\
\hline
DNABERT-2 & {\cellcolor[HTML]{DC5D4A}} \color[HTML]{F1F1F1} 56.76 & {\cellcolor[HTML]{B40426}} \color[HTML]{F1F1F1} 86.77 & {\cellcolor[HTML]{D0473D}} \color[HTML]{F1F1F1} 79.32 & {\cellcolor[HTML]{EE8468}} \color[HTML]{F1F1F1} 66.47 & {\cellcolor[HTML]{B40426}} \color[HTML]{F1F1F1} \textbf{52.66} \\
\hline
U-Net & {\cellcolor[HTML]{EAD4C8}} \color[HTML]{000000} 43.43 & {\cellcolor[HTML]{F5A081}} \color[HTML]{000000} 79.48 & {\cellcolor[HTML]{E67259}} \color[HTML]{F1F1F1} 73.89 & {\cellcolor[HTML]{F7B194}} \color[HTML]{000000} 57.11 & {\cellcolor[HTML]{F7B89C}} \color[HTML]{000000} 41.23 \\
\hline
U-Net-BPE & {\cellcolor[HTML]{3B4CC0}} \color[HTML]{F1F1F1} 20.67 & {\cellcolor[HTML]{3B4CC0}} \color[HTML]{F1F1F1} 59.41 & {\cellcolor[HTML]{3B4CC0}} \color[HTML]{F1F1F1} 17.97 & {\cellcolor[HTML]{3B4CC0}} \color[HTML]{F1F1F1} 1.56 & {\cellcolor[HTML]{455CCE}} \color[HTML]{F1F1F1} 19.56 \\
\hline
ConvNova & {\cellcolor[HTML]{C53334}} \color[HTML]{F1F1F1} 59.53 & {\cellcolor[HTML]{E97A5F}} \color[HTML]{F1F1F1} 81.88 & {\cellcolor[HTML]{B40426}} \color[HTML]{F1F1F1} 84.79 & {\cellcolor[HTML]{EE8669}} \color[HTML]{F1F1F1} 65.90 & {\cellcolor[HTML]{E97A5F}} \color[HTML]{F1F1F1} 46.61 \\
\hline
ConvNova-BPE & {\cellcolor[HTML]{6B8DF0}} \color[HTML]{F1F1F1} 26.91 & {\cellcolor[HTML]{A9C6FD}} \color[HTML]{000000} 68.43 & {\cellcolor[HTML]{D65244}} \color[HTML]{F1F1F1} 78.15 & {\cellcolor[HTML]{DFDBD9}} \color[HTML]{000000} 42.89 & {\cellcolor[HTML]{5673E0}} \color[HTML]{F1F1F1} 21.48 \\
\hline
\end{tabular}
\end{adjustbox}
\vspace{0.1cm}
\caption{MCC scores ($\times 100$) on the five mouse transcription factor binding datasets for DNABERT-2, U-Net, and ConvNova, comparing standard nucleotide-level tokenization with Byte Pair Encoding (BPE). Higher values indicate better performance.}
\end{table}

%% file: results_bpe/prom_table.tex
\begin{table}[!h]
\label{tab:prom_mcc_scores}
\centering 
\begin{adjustbox}{max width=\textwidth} 
\begin{tabular}{|l|c|c|c|c|c|c|}
\hline
\diagbox[width=12em]{Model}{Dataset} & prom\_300\_all & prom\_300\_notata & prom\_300\_tata & prom\_core\_all & prom\_core\_notata & prom\_core\_tata \\
\hline
DNABERT-2 & {\cellcolor[HTML]{CF453C}} \color[HTML]{F1F1F1} 86.77 & {\cellcolor[HTML]{B40426}} \color[HTML]{F1F1F1} 94.27 & {\cellcolor[HTML]{EC7F63}} \color[HTML]{F1F1F1} 71.59 & {\cellcolor[HTML]{DA5A49}} \color[HTML]{F1F1F1} 69.37 & {\cellcolor[HTML]{EC8165}} \color[HTML]{F1F1F1} 68.04 & {\cellcolor[HTML]{DF634E}} \color[HTML]{F1F1F1} 74.17 \\
\hline
U-Net & {\cellcolor[HTML]{C32E31}} \color[HTML]{F1F1F1} 87.73 & {\cellcolor[HTML]{DF634E}} \color[HTML]{F1F1F1} 92.24 & {\cellcolor[HTML]{B40426}} \color[HTML]{F1F1F1} 82.66 & {\cellcolor[HTML]{C83836}} \color[HTML]{F1F1F1} 71.04 & {\cellcolor[HTML]{BD1F2D}} \color[HTML]{F1F1F1} 71.59 & {\cellcolor[HTML]{B50927}} \color[HTML]{F1F1F1} 84.91 \\
\hline
U-Net-BPE & {\cellcolor[HTML]{3C4EC2}} \color[HTML]{F1F1F1} 59.30 & {\cellcolor[HTML]{3B4CC0}} \color[HTML]{F1F1F1} 78.72 & {\cellcolor[HTML]{4961D2}} \color[HTML]{F1F1F1} 27.20 & {\cellcolor[HTML]{3F53C6}} \color[HTML]{F1F1F1} 44.17 & {\cellcolor[HTML]{3B4CC0}} \color[HTML]{F1F1F1} 50.97 & {\cellcolor[HTML]{93B5FE}} \color[HTML]{000000} 22.44 \\
\hline
ConvNova & {\cellcolor[HTML]{F4987A}} \color[HTML]{000000} 81.60 & {\cellcolor[HTML]{E0654F}} \color[HTML]{F1F1F1} 92.20 & {\cellcolor[HTML]{EDD2C3}} \color[HTML]{000000} 57.51 & {\cellcolor[HTML]{DF634E}} \color[HTML]{F1F1F1} 68.87 & {\cellcolor[HTML]{DE614D}} \color[HTML]{F1F1F1} 69.39 & {\cellcolor[HTML]{D75445}} \color[HTML]{F1F1F1} 76.41 \\
\hline
ConvNova-BPE & {\cellcolor[HTML]{6485EC}} \color[HTML]{F1F1F1} 63.19 & {\cellcolor[HTML]{A7C5FE}} \color[HTML]{000000} 83.80 & {\cellcolor[HTML]{6F92F3}} \color[HTML]{F1F1F1} 34.05 & {\cellcolor[HTML]{6F92F3}} \color[HTML]{F1F1F1} 48.39 & {\cellcolor[HTML]{A5C3FE}} \color[HTML]{000000} 57.65 & {\cellcolor[HTML]{ECD3C5}} \color[HTML]{000000} 48.12 \\
\hline
\end{tabular}
\end{adjustbox}
\vspace{0.1cm}
\caption{MCC scores ($\times 100$) on the six promoter identification datasets for DNABERT-2, U-Net, and ConvNova, comparing standard nucleotide-level tokenization with Byte Pair Encoding (BPE). The benchmark includes both promoter and core-promoter prediction tasks, with and without TATA-box stratification. Higher values indicate better performance.}
\end{table}

%% file: results_bpe/splice_table.tex
\begin{table}[!h]
\label{tab:splice_mcc_scores_bpe}
\centering 
\begin{adjustbox}{max width=\textwidth} 
\begin{tabular}{|l|c|c|c|}
\hline
\diagbox[width=12em]{Model}{Dataset} & reconstructed \\
\hline
DNABERT-2 & {\cellcolor[HTML]{C32E31}} \color[HTML]{F1F1F1} 84.99 \\
\hline
U-Net & {\cellcolor[HTML]{BB1B2C}} \color[HTML]{F1F1F1} 86.16 \\
\hline
U-Net-BPE & {\cellcolor[HTML]{3B4CC0}} \color[HTML]{F1F1F1} 22.91 \\
\hline
ConvNova & {\cellcolor[HTML]{C43032}} \color[HTML]{F1F1F1} 84.64 \\
\hline
ConvNova-BPE & {\cellcolor[HTML]{F7B99E}} \color[HTML]{000000} 65.66 \\
\hline
\end{tabular}
\end{adjustbox}
\vspace{0.1cm}
\caption{MCC scores ($\times 100$) on the splice-site benchmark for DNABERT-2, U-Net, and ConvNova, comparing standard nucleotide-level tokenization with Byte Pair Encoding (BPE). Higher values indicate better performance.}
\label{tab:splice_mcc_scores_bpe}
\end{table}

%% file: results_bpe/tf_table.tex
\begin{table}[!h]

\centering 
\begin{adjustbox}{max width=\textwidth} 
\begin{tabular}{|l|c|c|c|c|c|}
\hline
\diagbox[width=12em]{Model}{Dataset} & 0 & 1 & 2 & 3 & 4 \\
\hline
DNABERT-2 & {\cellcolor[HTML]{B40426}} \color[HTML]{F1F1F1} 71.99 & {\cellcolor[HTML]{B40426}} \color[HTML]{F1F1F1} 76.06 & {\cellcolor[HTML]{B40426}} \color[HTML]{F1F1F1} 66.52 & {\cellcolor[HTML]{B40426}} \color[HTML]{F1F1F1} 58.54 & {\cellcolor[HTML]{B40426}} \color[HTML]{F1F1F1} \textbf{77.43} \\
\hline
U-Net & {\cellcolor[HTML]{97B8FF}} \color[HTML]{000000} 43.44 & {\cellcolor[HTML]{A2C1FF}} \color[HTML]{000000} 45.55 & {\cellcolor[HTML]{F7BCA1}} \color[HTML]{000000} 50.44 & {\cellcolor[HTML]{B6CEFA}} \color[HTML]{000000} 26.88 & {\cellcolor[HTML]{C1D4F4}} \color[HTML]{000000} 44.42 \\
\hline
U-Net-BPE & {\cellcolor[HTML]{4257C9}} \color[HTML]{F1F1F1} 33.28 & {\cellcolor[HTML]{3B4CC0}} \color[HTML]{F1F1F1} 31.78 & {\cellcolor[HTML]{3B4CC0}} \color[HTML]{F1F1F1} 19.92 & {\cellcolor[HTML]{445ACC}} \color[HTML]{F1F1F1} 10.02 & {\cellcolor[HTML]{4B64D5}} \color[HTML]{F1F1F1} 25.27 \\
\hline
ConvNova & {\cellcolor[HTML]{BFD3F6}} \color[HTML]{000000} 48.05 & {\cellcolor[HTML]{A1C0FF}} \color[HTML]{000000} 45.40 & {\cellcolor[HTML]{F7A98B}} \color[HTML]{000000} 52.92 & {\cellcolor[HTML]{B9D0F9}} \color[HTML]{000000} 27.31 & {\cellcolor[HTML]{D5DBE5}} \color[HTML]{000000} 48.15 \\
\hline
ConvNova-BPE & {\cellcolor[HTML]{A6C4FE}} \color[HTML]{000000} 45.13 & {\cellcolor[HTML]{4B64D5}} \color[HTML]{F1F1F1} 34.34 & {\cellcolor[HTML]{90B2FE}} \color[HTML]{000000} 32.10 & {\cellcolor[HTML]{A1C0FF}} \color[HTML]{000000} 23.69 & {\cellcolor[HTML]{3B4CC0}} \color[HTML]{F1F1F1} 22.08 \\
\hline
\end{tabular}
\end{adjustbox}
\vspace{0.1cm}
\caption{MCC scores ($\times 100$) on the five transcription factor binding benchmark datasets for DNABERT-2, U-Net, and ConvNova, comparing standard nucleotide-level tokenization with Byte Pair Encoding (BPE). Higher values indicate better performance.}
\label{tab:tf_mcc_scores}
\end{table}

%% file: results_bpe/virus_table.tex
\begin{table}[!h]
\label{tab:virus_mcc_scores}
\centering 
\begin{adjustbox}{max width=\textwidth} 
\begin{tabular}{|l|c|c|c|}
\hline
\diagbox[width=12em]{Model}{Dataset} & covid \\
\hline
DNABERT-2 & {\cellcolor[HTML]{B8122A}} \color[HTML]{F1F1F1} 71.02 \\
\hline
U-Net & {\cellcolor[HTML]{6C8FF1}} \color[HTML]{F1F1F1} 65.99 \\
\hline
U-Net-BPE & {\cellcolor[HTML]{DDDCDC}} \color[HTML]{000000} 68.07 \\
\hline
ConvNova & {\cellcolor[HTML]{F2CAB5}} \color[HTML]{000000} 68.68 \\
\hline
ConvNova-BPE & {\cellcolor[HTML]{BB1B2C}} \color[HTML]{F1F1F1} 70.98 \\
\hline
\end{tabular}
\end{adjustbox}
\vspace{0.1cm}
\caption{MCC scores ($\times 100$) on the virus classification benchmark for DNABERT-2, U-Net, and ConvNova, comparing standard nucleotide-level tokenization with Byte Pair Encoding (BPE). Higher values indicate better performance.}
\end{table}

%% file: results_pretraining/EMP_table.tex
\begin{table}[!h]

\centering 
\begin{adjustbox}{max width=\textwidth} 
\begin{tabular}{|l|c|c|c|c|c|c|c|c|c|c|}
\hline
\diagbox[width=12em]{Model}{Dataset}  & H3 & H3K14ac & H3K36me3 & H3K4me1 & H3K4me2 & H3K4me3 & H3K79me3 & H3K9ac & H4 & H4ac \\ \hline
DNABERT-2 & {\cellcolor[HTML]{B8122A}} \color[HTML]{F1F1F1} 78.27 & {\cellcolor[HTML]{EF886B}} \color[HTML]{F1F1F1} 52.57 & {\cellcolor[HTML]{B40426}} \color[HTML]{F1F1F1} 56.88 & {\cellcolor[HTML]{B40426}} \color[HTML]{F1F1F1} 50.52 & {\cellcolor[HTML]{CBD8EE}} \color[HTML]{000000} 31.13 & {\cellcolor[HTML]{F5C2AA}} \color[HTML]{000000} 36.27 & {\cellcolor[HTML]{B40426}} \color[HTML]{F1F1F1} 67.39 & {\cellcolor[HTML]{CF453C}} \color[HTML]{F1F1F1} 55.63 & {\cellcolor[HTML]{B40426}} \color[HTML]{F1F1F1} 80.71 & {\cellcolor[HTML]{EC7F63}} \color[HTML]{F1F1F1} 50.43 \\
\hline
DNABERT-2 (no pretraining) & {\cellcolor[HTML]{A6C4FE}} \color[HTML]{000000} 60.57 & {\cellcolor[HTML]{8CAFFE}} \color[HTML]{000000} 32.71 & {\cellcolor[HTML]{799CF8}} \color[HTML]{F1F1F1} 40.67 & {\cellcolor[HTML]{5A78E4}} \color[HTML]{F1F1F1} 25.93 & {\cellcolor[HTML]{9FBFFF}} \color[HTML]{000000} 27.91 & {\cellcolor[HTML]{8DB0FE}} \color[HTML]{000000} 23.82 & {\cellcolor[HTML]{ADC9FD}} \color[HTML]{000000} 55.26 & {\cellcolor[HTML]{A5C3FE}} \color[HTML]{000000} 45.77 & {\cellcolor[HTML]{E3D9D3}} \color[HTML]{000000} 68.09 & {\cellcolor[HTML]{CDD9EC}} \color[HTML]{000000} 41.98 \\
\hline
U-Net (pretrained) & {\cellcolor[HTML]{F49A7B}} \color[HTML]{000000} 71.88 & {\cellcolor[HTML]{A5C3FE}} \color[HTML]{000000} 35.22 & {\cellcolor[HTML]{DBDCDE}} \color[HTML]{000000} 46.74 & {\cellcolor[HTML]{D3DBE7}} \color[HTML]{000000} 35.79 & {\cellcolor[HTML]{82A6FB}} \color[HTML]{F1F1F1} 26.04 & {\cellcolor[HTML]{94B6FF}} \color[HTML]{000000} 24.50 & {\cellcolor[HTML]{F6A586}} \color[HTML]{000000} 62.26 & {\cellcolor[HTML]{F29072}} \color[HTML]{F1F1F1} 53.14 & {\cellcolor[HTML]{DA5A49}} \color[HTML]{F1F1F1} 77.61 & {\cellcolor[HTML]{8CAFFE}} \color[HTML]{000000} 37.55 \\
\hline
U-Net (pretrained-human only) & {\cellcolor[HTML]{F59C7D}} \color[HTML]{000000} 71.81 & {\cellcolor[HTML]{A2C1FF}} \color[HTML]{000000} 35.01 & {\cellcolor[HTML]{DFDBD9}} \color[HTML]{000000} 47.05 & {\cellcolor[HTML]{C1D4F4}} \color[HTML]{000000} 34.12 & {\cellcolor[HTML]{8BADFD}} \color[HTML]{000000} 26.58 & {\cellcolor[HTML]{A1C0FF}} \color[HTML]{000000} 25.66 & {\cellcolor[HTML]{F7AD90}} \color[HTML]{000000} 61.83 & {\cellcolor[HTML]{E67259}} \color[HTML]{F1F1F1} 54.19 & {\cellcolor[HTML]{F18F71}} \color[HTML]{F1F1F1} 74.75 & {\cellcolor[HTML]{8CAFFE}} \color[HTML]{000000} 37.54 \\
\hline
U-Net (no pretraining) & {\cellcolor[HTML]{F6A283}} \color[HTML]{000000} 71.40 & {\cellcolor[HTML]{B5CDFA}} \color[HTML]{000000} 37.00 & {\cellcolor[HTML]{D3DBE7}} \color[HTML]{000000} 46.07 & {\cellcolor[HTML]{B6CEFA}} \color[HTML]{000000} 33.19 & {\cellcolor[HTML]{B1CBFC}} \color[HTML]{000000} 29.13 & {\cellcolor[HTML]{B2CCFB}} \color[HTML]{000000} 27.24 & {\cellcolor[HTML]{F39475}} \color[HTML]{000000} 63.07 & {\cellcolor[HTML]{F7B396}} \color[HTML]{000000} 51.65 & {\cellcolor[HTML]{F6A283}} \color[HTML]{000000} 73.50 & {\cellcolor[HTML]{9FBFFF}} \color[HTML]{000000} 38.80 \\
\hline
ConvNova (pretrained) & {\cellcolor[HTML]{EA7B60}} \color[HTML]{F1F1F1} 73.77 & {\cellcolor[HTML]{F7AA8C}} \color[HTML]{000000} 49.42 & {\cellcolor[HTML]{D75445}} \color[HTML]{F1F1F1} 54.83 & {\cellcolor[HTML]{F4987A}} \color[HTML]{000000} 43.67 & {\cellcolor[HTML]{DADCE0}} \color[HTML]{000000} 32.41 & {\cellcolor[HTML]{BE242E}} \color[HTML]{F1F1F1} 47.42 & {\cellcolor[HTML]{C43032}} \color[HTML]{F1F1F1} 66.54 & {\cellcolor[HTML]{C83836}} \color[HTML]{F1F1F1} 55.92 & {\cellcolor[HTML]{E57058}} \color[HTML]{F1F1F1} 76.56 & {\cellcolor[HTML]{F08B6E}} \color[HTML]{F1F1F1} 49.80 \\
\hline
ConvNova (pretrained-human only) & {\cellcolor[HTML]{B40426}} \color[HTML]{F1F1F1} 78.62 & {\cellcolor[HTML]{F49A7B}} \color[HTML]{000000} 51.07 & {\cellcolor[HTML]{EE8669}} \color[HTML]{F1F1F1} 52.78 & {\cellcolor[HTML]{EDD2C3}} \color[HTML]{000000} 38.66 & {\cellcolor[HTML]{F7B396}} \color[HTML]{000000} 36.95 & {\cellcolor[HTML]{ED8366}} \color[HTML]{F1F1F1} 41.89 & {\cellcolor[HTML]{F7AD90}} \color[HTML]{000000} 61.82 & {\cellcolor[HTML]{B6CEFA}} \color[HTML]{000000} 46.54 & {\cellcolor[HTML]{F5A081}} \color[HTML]{000000} 73.65 & {\cellcolor[HTML]{E3D9D3}} \color[HTML]{000000} 43.94 \\
\hline
ConvNova (no pretraining) & {\cellcolor[HTML]{EA7B60}} \color[HTML]{F1F1F1} 73.77 & {\cellcolor[HTML]{CC403A}} \color[HTML]{F1F1F1} 57.72 & {\cellcolor[HTML]{D75445}} \color[HTML]{F1F1F1} 54.83 & {\cellcolor[HTML]{F7AA8C}} \color[HTML]{000000} 42.37 & {\cellcolor[HTML]{DADCE0}} \color[HTML]{000000} 32.41 & {\cellcolor[HTML]{F59C7D}} \color[HTML]{000000} 39.96 & {\cellcolor[HTML]{C0282F}} \color[HTML]{F1F1F1} 66.81 & {\cellcolor[HTML]{CD423B}} \color[HTML]{F1F1F1} 55.70 & {\cellcolor[HTML]{E67259}} \color[HTML]{F1F1F1} 76.42 & {\cellcolor[HTML]{F08B6E}} \color[HTML]{F1F1F1} 49.80 \\
\hline
\end{tabular}
\end{adjustbox}

\vspace{0.1cm}
\caption{MCC scores ($\times 100$) on the EMP benchmark under different pretraining regimes. DNABERT-2 is reported as the pretrained reference model and as a randomly initialized model, while U-Net and ConvNova are compared across multi-species pretraining, human-only pretraining, and no pretraining. The EMP benchmark comprises 10 epigenetic mark prediction datasets. Higher values indicate better performance.}
\label{tab:EMP_mcc_scores_pt}

\end{table}

%% file: results_pretraining/mouse_table.tex
\begin{table}[!h]
\label{tab:mouse_mcc_scores_}
\centering 
\begin{adjustbox}{max width=\textwidth} 
\begin{tabular}{|l|c|c|c|c|c|c|c|c|c|c|}
\hline
\diagbox[width=12em]{Model}{Dataset} & 0 & 1 & 2 & 3 & 4 \\
\hline
DNABERT-2& {\cellcolor[HTML]{DC5D4A}} \color[HTML]{F1F1F1} 56.76 & {\cellcolor[HTML]{B40426}} \color[HTML]{F1F1F1} 86.77 & {\cellcolor[HTML]{D0473D}} \color[HTML]{F1F1F1} 79.32 & {\cellcolor[HTML]{EE8468}} \color[HTML]{F1F1F1} 66.47 & {\cellcolor[HTML]{B40426}} \color[HTML]{F1F1F1} \textbf{52.66} \\
\hline
DNABERT-2 (no pretraining) & {\cellcolor[HTML]{C0D4F5}} \color[HTML]{000000} 37.03 & {\cellcolor[HTML]{D7DCE3}} \color[HTML]{000000} 72.47 & {\cellcolor[HTML]{EC7F63}} \color[HTML]{F1F1F1} 72.17 & {\cellcolor[HTML]{DDDCDC}} \color[HTML]{000000} 42.17 & {\cellcolor[HTML]{8FB1FE}} \color[HTML]{000000} 27.13 \\
\hline
U-Net (pretrained) & {\cellcolor[HTML]{EAD4C8}} \color[HTML]{000000} 43.43 & {\cellcolor[HTML]{F5A081}} \color[HTML]{000000} 79.48 & {\cellcolor[HTML]{E67259}} \color[HTML]{F1F1F1} 73.89 & {\cellcolor[HTML]{F7B194}} \color[HTML]{000000} 57.11 & {\cellcolor[HTML]{F7B89C}} \color[HTML]{000000} 41.23 \\
\hline
U-Net (pretrained-human only) & {\cellcolor[HTML]{E7745B}} \color[HTML]{F1F1F1} 54.78 & {\cellcolor[HTML]{F7AF91}} \color[HTML]{000000} 78.33 & {\cellcolor[HTML]{F18F71}} \color[HTML]{F1F1F1} 69.84 & {\cellcolor[HTML]{F59F80}} \color[HTML]{000000} 61.37 & {\cellcolor[HTML]{F7B194}} \color[HTML]{000000} 41.84 \\
\hline
U-Net (no pretraining) & {\cellcolor[HTML]{F7B79B}} \color[HTML]{000000} 48.14 & {\cellcolor[HTML]{EE8669}} \color[HTML]{F1F1F1} 81.20 & {\cellcolor[HTML]{E97A5F}} \color[HTML]{F1F1F1} 72.87 & {\cellcolor[HTML]{CEDAEB}} \color[HTML]{000000} 37.96 & {\cellcolor[HTML]{F7B99E}} \color[HTML]{000000} 41.00 \\
\hline
ConvNova (pretrained)  & {\cellcolor[HTML]{C53334}} \color[HTML]{F1F1F1} 59.53 & {\cellcolor[HTML]{E97A5F}} \color[HTML]{F1F1F1} 81.88 & {\cellcolor[HTML]{B40426}} \color[HTML]{F1F1F1} 84.79 & {\cellcolor[HTML]{EE8669}} \color[HTML]{F1F1F1} 65.90 & {\cellcolor[HTML]{E97A5F}} \color[HTML]{F1F1F1} 46.61 \\
\hline
ConvNova (pretrained-human only) & {\cellcolor[HTML]{C43032}} \color[HTML]{F1F1F1} 59.80 & {\cellcolor[HTML]{EE8669}} \color[HTML]{F1F1F1} 81.13 & {\cellcolor[HTML]{DD5F4B}} \color[HTML]{F1F1F1} 76.63 & {\cellcolor[HTML]{B40426}} \color[HTML]{F1F1F1} 82.62 & {\cellcolor[HTML]{F6A385}} \color[HTML]{000000} 43.21 \\
\hline
ConvNova (no pretraining) & {\cellcolor[HTML]{D75445}} \color[HTML]{F1F1F1} 57.36 & {\cellcolor[HTML]{F59C7D}} \color[HTML]{000000} 79.77 & {\cellcolor[HTML]{C12B30}} \color[HTML]{F1F1F1} 82.19 & {\cellcolor[HTML]{F08B6E}} \color[HTML]{F1F1F1} 64.94 & {\cellcolor[HTML]{E46E56}} \color[HTML]{F1F1F1} 47.30 \\
\hline
\end{tabular}
\end{adjustbox}
\vspace{0.1cm}
\caption{MCC scores ($\times 100$) on the five mouse transcription factor binding datasets under different pretraining regimes. DNABERT-2 is reported as the pretrained reference model and as a randomly initialized model, while U-Net and ConvNova are compared across multi-species pretraining, human-only pretraining, and no pretraining. Higher values indicate better performance.}
\end{table}

%% file: results_pretraining/prom_table.tex
\begin{table}[!h]
\label{tab:prom_mcc_scores_}
\centering 
\begin{adjustbox}{max width=\textwidth} 
\begin{tabular}{|l|c|c|c|c|c|c|c|c|c|c|}
\hline
\diagbox[width=12em]{Model}{Dataset} & prom\_300\_all & prom\_300\_notata & prom\_300\_tata & prom\_core\_all & prom\_core\_notata & prom\_core\_tata \\
\hline
DNABERT-2 & {\cellcolor[HTML]{CF453C}} \color[HTML]{F1F1F1} 86.77 & {\cellcolor[HTML]{B40426}} \color[HTML]{F1F1F1} 94.27 & {\cellcolor[HTML]{EC7F63}} \color[HTML]{F1F1F1} 71.59 & {\cellcolor[HTML]{DA5A49}} \color[HTML]{F1F1F1} 69.37 & {\cellcolor[HTML]{EC8165}} \color[HTML]{F1F1F1} 68.04 & {\cellcolor[HTML]{DF634E}} \color[HTML]{F1F1F1} 74.17 \\
\hline
DNABERT-2 (no pretraining) & {\cellcolor[HTML]{7396F5}} \color[HTML]{F1F1F1} 64.37 & {\cellcolor[HTML]{A9C6FD}} \color[HTML]{000000} 83.86 & {\cellcolor[HTML]{5E7DE7}} \color[HTML]{F1F1F1} 31.18 & {\cellcolor[HTML]{7396F5}} \color[HTML]{F1F1F1} 48.80 & {\cellcolor[HTML]{C4D5F3}} \color[HTML]{000000} 59.68 & {\cellcolor[HTML]{CEDAEB}} \color[HTML]{000000} 37.85 \\
\hline
U-Net (pretrained) & {\cellcolor[HTML]{C32E31}} \color[HTML]{F1F1F1} 87.73 & {\cellcolor[HTML]{DF634E}} \color[HTML]{F1F1F1} 92.24 & {\cellcolor[HTML]{B40426}} \color[HTML]{F1F1F1} 82.66 & {\cellcolor[HTML]{C83836}} \color[HTML]{F1F1F1} 71.04 & {\cellcolor[HTML]{BD1F2D}} \color[HTML]{F1F1F1} 71.59 & {\cellcolor[HTML]{B50927}} \color[HTML]{F1F1F1} 84.91 \\
\hline
U-Net (pretrained-human only) & {\cellcolor[HTML]{BB1B2C}} \color[HTML]{F1F1F1} 88.34 & {\cellcolor[HTML]{E57058}} \color[HTML]{F1F1F1} 91.79 & {\cellcolor[HTML]{C32E31}} \color[HTML]{F1F1F1} 80.34 & {\cellcolor[HTML]{C83836}} \color[HTML]{F1F1F1} 70.99 & {\cellcolor[HTML]{B70D28}} \color[HTML]{F1F1F1} 71.95 & {\cellcolor[HTML]{B40426}} \color[HTML]{F1F1F1} \textbf{85.36} \\
\hline
U-Net (no pretraining) & {\cellcolor[HTML]{B40426}} \color[HTML]{F1F1F1} 89.01 & {\cellcolor[HTML]{EA7B60}} \color[HTML]{F1F1F1} 91.44 & {\cellcolor[HTML]{C43032}} \color[HTML]{F1F1F1} 79.98 & {\cellcolor[HTML]{D44E41}} \color[HTML]{F1F1F1} 69.96 & {\cellcolor[HTML]{B40426}} \color[HTML]{F1F1F1} 72.12 & {\cellcolor[HTML]{B70D28}} \color[HTML]{F1F1F1} 84.68 \\
\hline
ConvNova (pretrained) & {\cellcolor[HTML]{F4987A}} \color[HTML]{000000} 81.60 & {\cellcolor[HTML]{E0654F}} \color[HTML]{F1F1F1} 92.20 & {\cellcolor[HTML]{EDD2C3}} \color[HTML]{000000} 57.51 & {\cellcolor[HTML]{DF634E}} \color[HTML]{F1F1F1} 68.87 & {\cellcolor[HTML]{DE614D}} \color[HTML]{F1F1F1} 69.39 & {\cellcolor[HTML]{D75445}} \color[HTML]{F1F1F1} 76.41 \\
\hline
ConvNova (pretrained-human only) & {\cellcolor[HTML]{ED8366}} \color[HTML]{F1F1F1} 83.14 & {\cellcolor[HTML]{E9785D}} \color[HTML]{F1F1F1} 91.59 & {\cellcolor[HTML]{E4D9D2}} \color[HTML]{000000} 55.32 & {\cellcolor[HTML]{B40426}} \color[HTML]{F1F1F1} 72.67 & {\cellcolor[HTML]{DD5F4B}} \color[HTML]{F1F1F1} 69.50 & {\cellcolor[HTML]{D0473D}} \color[HTML]{F1F1F1} 78.44 \\
\hline
ConvNova (no pretraining) & {\cellcolor[HTML]{F7A98B}} \color[HTML]{000000} 80.26 & {\cellcolor[HTML]{DF634E}} \color[HTML]{F1F1F1} 92.23 & {\cellcolor[HTML]{E8D6CC}} \color[HTML]{000000} 56.23 & {\cellcolor[HTML]{DC5D4A}} \color[HTML]{F1F1F1} 69.21 & {\cellcolor[HTML]{D55042}} \color[HTML]{F1F1F1} 70.07 & {\cellcolor[HTML]{D75445}} \color[HTML]{F1F1F1} 76.52 \\
\hline
\end{tabular}
\end{adjustbox}
\vspace{0.1cm}
\caption{MCC scores ($\times 100$) on the six promoter identification datasets under different pretraining regimes. DNABERT-2 is reported as the pretrained reference model and as a randomly initialized model, while U-Net and ConvNova are compared across multi-species pretraining, human-only pretraining, and no pretraining. The benchmark includes both promoter and core-promoter tasks, with and without TATA-box stratification. Higher values indicate better performance.}
\end{table}

%% file: results_pretraining/splice_table.tex
\begin{table}[!h]

\centering 
\begin{adjustbox}{max width=\textwidth} 
\begin{tabular}{|l|c|c|c|c|c|c|c|c|c|c|}
\hline
\diagbox[width=12em]{Model}{Dataset} & reconstructed \\
\hline
DNABERT-2 & {\cellcolor[HTML]{C32E31}} \color[HTML]{F1F1F1} 84.99 \\
\hline
DNABERT-2 (no pretraining) & {\cellcolor[HTML]{DEDCDB}} \color[HTML]{000000} 55.53 \\
\hline
U-Net (pretrained) & {\cellcolor[HTML]{BB1B2C}} \color[HTML]{F1F1F1} 86.16 \\
\hline
U-Net (pretrained-human only) & {\cellcolor[HTML]{B40426}} \color[HTML]{F1F1F1} \textbf{87.54} \\
\hline
U-Net (no pretraining) & {\cellcolor[HTML]{C32E31}} \color[HTML]{F1F1F1} 84.77 \\
\hline
ConvNova (pretrained) & {\cellcolor[HTML]{C43032}} \color[HTML]{F1F1F1} 84.64 \\
\hline
ConvNova (pretrained-human only) & {\cellcolor[HTML]{B70D28}} \color[HTML]{F1F1F1} 87.03 \\
\hline
ConvNova (no pretraining) & {\cellcolor[HTML]{BA162B}} \color[HTML]{F1F1F1} 86.44 \\
\hline
\end{tabular}
\end{adjustbox}
\vspace{0.1cm}
\caption{MCC scores ($\times 100$) on the splice-site benchmark under different pretraining regimes. DNABERT-2 is reported as the pretrained reference model and as a randomly initialized model, while U-Net and ConvNova are compared across multi-species pretraining, human-only pretraining, and no pretraining. Higher values indicate better performance.}
\label{tab:splice_mcc_scores_pt}
\end{table}

%% file: results_pretraining/tf_table.tex
\begin{table}[h]

\centering 
\begin{adjustbox}{max width=\textwidth} 
\begin{tabular}{|l|c|c|c|c|c|}
\hline
\diagbox[width=12em]{Model}{Dataset} & 0 & 1 & 2 & 3 & 4 \\
\hline
DNABERT-2 & {\cellcolor[HTML]{B40426}} \color[HTML]{F1F1F1} 71.99 & {\cellcolor[HTML]{B40426}} \color[HTML]{F1F1F1} 76.06 & {\cellcolor[HTML]{B40426}} \color[HTML]{F1F1F1} 66.52 & {\cellcolor[HTML]{B40426}} \color[HTML]{F1F1F1} 58.54 & {\cellcolor[HTML]{B40426}} \color[HTML]{F1F1F1} \textbf{77.43} \\
\hline
DNABERT-2 (no pretraining) & {\cellcolor[HTML]{8FB1FE}} \color[HTML]{000000} 42.48 & {\cellcolor[HTML]{6384EB}} \color[HTML]{F1F1F1} 37.53 & {\cellcolor[HTML]{688AEF}} \color[HTML]{F1F1F1} 26.73 & {\cellcolor[HTML]{93B5FE}} \color[HTML]{000000} 21.78 & {\cellcolor[HTML]{8CAFFE}} \color[HTML]{000000} 35.88 \\
\hline
U-Net (pretrained) & {\cellcolor[HTML]{97B8FF}} \color[HTML]{000000} 43.44 & {\cellcolor[HTML]{A2C1FF}} \color[HTML]{000000} 45.55 & {\cellcolor[HTML]{F7BCA1}} \color[HTML]{000000} 50.44 & {\cellcolor[HTML]{B6CEFA}} \color[HTML]{000000} 26.88 & {\cellcolor[HTML]{C1D4F4}} \color[HTML]{000000} 44.42 \\
\hline
U-Net (pretrained-human only) & {\cellcolor[HTML]{93B5FE}} \color[HTML]{000000} 43.01 & {\cellcolor[HTML]{8BADFD}} \color[HTML]{000000} 42.59 & {\cellcolor[HTML]{F4C6AF}} \color[HTML]{000000} 48.74 & {\cellcolor[HTML]{B2CCFB}} \color[HTML]{000000} 26.23 & {\cellcolor[HTML]{BED2F6}} \color[HTML]{000000} 43.90 \\
\hline
U-Net (no pretraining) & {\cellcolor[HTML]{9EBEFF}} \color[HTML]{000000} 44.18 & {\cellcolor[HTML]{C5D6F2}} \color[HTML]{000000} 50.17 & {\cellcolor[HTML]{EF886B}} \color[HTML]{F1F1F1} 56.81 & {\cellcolor[HTML]{B1CBFC}} \color[HTML]{000000} 26.19 & {\cellcolor[HTML]{B3CDFB}} \color[HTML]{000000} 42.08 \\
\hline
ConvNova (pretrained) & {\cellcolor[HTML]{BFD3F6}} \color[HTML]{000000} 48.05 & {\cellcolor[HTML]{A1C0FF}} \color[HTML]{000000} 45.40 & {\cellcolor[HTML]{F7A98B}} \color[HTML]{000000} 52.92 & {\cellcolor[HTML]{B9D0F9}} \color[HTML]{000000} 27.31 & {\cellcolor[HTML]{D5DBE5}} \color[HTML]{000000} 48.15 \\
\hline
ConvNova (pretrained-human only) & {\cellcolor[HTML]{7597F6}} \color[HTML]{F1F1F1} 39.56 & {\cellcolor[HTML]{D2DBE8}} \color[HTML]{000000} 52.03 & {\cellcolor[HTML]{F1CCB8}} \color[HTML]{000000} 47.61 & {\cellcolor[HTML]{D1DAE9}} \color[HTML]{000000} 31.29 & {\cellcolor[HTML]{E0DBD8}} \color[HTML]{000000} 50.55 \\
\hline
ConvNova (no pretraining) & {\cellcolor[HTML]{AEC9FC}} \color[HTML]{000000} 45.98 & {\cellcolor[HTML]{A5C3FE}} \color[HTML]{000000} 45.91 & {\cellcolor[HTML]{F59F80}} \color[HTML]{000000} 54.17 & {\cellcolor[HTML]{F6BFA6}} \color[HTML]{000000} 40.71 & {\cellcolor[HTML]{DBDCDE}} \color[HTML]{000000} 49.34 \\
\hline
\end{tabular}
\end{adjustbox}
\vspace{0.1cm}
\caption{MCC scores ($\times 100$) on the five transcription factor binding benchmark datasets under different pretraining regimes. DNABERT-2 is reported as the pretrained reference model and as a randomly initialized model, while U-Net and ConvNova are compared across multi-species pretraining, human-only pretraining, and no pretraining. Higher values indicate better performance.}
\label{tab:tf_mcc_scores_pt}
\end{table}

%% file: results_pretraining/virus_table.tex
\begin{table}[!h]

\centering 
\begin{adjustbox}{max width=\textwidth} 
\begin{tabular}{|l|c|c|}
\hline
\diagbox[width=12em]{Model}{Dataset} & covid \\
\hline
DNABERT-2 & {\cellcolor[HTML]{B8122A}} \color[HTML]{F1F1F1} 71.02 \\
\hline
DNABERT-2 (no pretraining) & {\cellcolor[HTML]{D6DCE4}} \color[HTML]{000000} 67.90 \\
\hline
U-Net (pretrained) & {\cellcolor[HTML]{6C8FF1}} \color[HTML]{F1F1F1} 65.99 \\
\hline
U-Net (pretrained-human only) & {\cellcolor[HTML]{94B6FF}} \color[HTML]{000000} 66.67 \\
\hline
U-Net (no pretraining) & {\cellcolor[HTML]{3B4CC0}} \color[HTML]{F1F1F1} 65.02 \\
\hline
ConvNova (pretrained) & {\cellcolor[HTML]{F2CAB5}} \color[HTML]{000000} 68.68 \\
\hline
ConvNova (pretrained-human only) & {\cellcolor[HTML]{D5DBE5}} \color[HTML]{000000} 67.87 \\
\hline
ConvNova (no pretraining) & {\cellcolor[HTML]{D5DBE5}} \color[HTML]{000000} 67.89 \\
\hline
\end{tabular}
\end{adjustbox}
\caption{MCC scores ($\times 100$) on the virus classification benchmark under different pretraining regimes. DNABERT-2 is reported as the pretrained reference model and as a randomly initialized model, while U-Net and ConvNova are compared across multi-species pretraining, human-only pretraining, and no pretraining. Higher values indicate better performance.}
\label{tab:virus_mcc_scores_}
\end{table}